\documentclass[12pt]{article}
\usepackage{cite}
\usepackage{amssymb}
\usepackage{epsfig}

\newcommand{\nn}{\nonumber}
\newcommand{\be}{\begin{equation}}                
\newcommand{\ee}{\end{equation}}        
\newcommand{\bea}{\begin{eqnarray}}               
\newcommand{\eea}{\end{eqnarray}}        
\newcommand{\BR}{{\rm BR}}        
\catcode`@=11 
\def\lsim{\mathrel{\mathpalette\@versim<}}
\def\gsim{\mathrel{\mathpalette\@versim>}}
 \def\@versim#1#2{\lower0.2ex\vbox{\baselineskip\z@skip\lineskip\z@skip
       \lineskiplimit\z@\ialign{$\m@th#1\hfil##$\crcr#2\crcr\sim\crcr}}}

\begin{document}
\pagestyle{empty}
\begin{flushright}
CERN-TH/2002-069\\
RM3-TH/02-4
\end{flushright}
\vspace*{15mm}

\begin{center}
\centerline{\LARGE\bf  A new model-independent way} 
\vskip 2mm
\centerline{\LARGE\bf  of extracting 
$|V_{ub}/V_{cb}|$  }
\vskip1truecm
\centerline{\large\bf Ugo Aglietti$^a$\footnote{Present address: Dip. di Fisica,
Universit\`a di Roma ``La Sapienza'', Piazzale Aldo Moro 2, I-00185,
Rome, Italy.}, Marco Ciuchini$^b$, Paolo
  Gambino$^a$}
\bigskip
\vspace{0.3cm}
$^a$ Theoretical Physics Division, CERN, \\[1pt]
CH-1211, Geneva 23, Switzerland \\[3pt]
$^b$ Dip. di Fisica, Universit\`a di Roma Tre and I.N.F.N. Sezione di Roma III, \\[1pt]
Via della Vasca Navale 84, I-00146, Rome, Italy   \\[3pt]
\vspace*{2cm} \textbf{Abstract} \\[0pt]
\end{center}
The ratio between the photon spectrum in $B\to X_s \gamma$
and the differential semileptonic rate wrt the hadronic variable 
$M_X/E_X$ is a short-distance quantity calculable in perturbation
theory  and independent of the Fermi motion of the $b$ quark in the $B$ meson. 
We present a NLO analysis of this ratio and show how it  can be used
to determine  $\vert V_{ub}/V_{cb}\vert$
independently of any model for the shape function.
We also discuss how this relation
can be used to test the validity of the shape-function theory on the data.
\vfill
\newpage
\setcounter{page}{1} \pagestyle{plain}

\section{Introduction}
The accurate determination of the parameters entering the Cabibbo--Kobayashi--Maskawa (CKM) matrix  $V$
is one of the main physics goals at beauty factories. The large number of produced $B$ mesons allows
for the study of many decay modes and the measurement of branching ratios (BRs) as small as $10^{-6}$.
Unfortunately, a precise extraction of the CKM parameters from the data is often hindered by uncertainties
on the hadronic parameters accounting for low-energy strong interaction effects.
There exist, however, observable quantities that are free, 
at least to some extent, from
hadronic uncertainties. They are obviously highly valued and 
have been intensively looked for in recent years.

In particular, it has been suggested many years ago that one can build a suitable ratio of the lepton-energy spectrum
in the semileptonic decay  $B\to X_u l \nu$ and the photon spectrum in
$B\to X_s\gamma$  from which the dominant non-perturbative
QCD effects due to the heavy-quark Fermi motion drop out~\cite{Bigi:1993ex,Neubert:1993um}.
This ratio can be used to extract the combination of CKM matrix elements $\vert V_{ub}/(V_{tb} V_{ts}^*)\vert$.
Theoretically, the cancellation of long-distance effects takes place for hadronic invariant masses as small as
$M_X^2\sim Q \Lambda_{QCD}$, where $M_X$ is the invariant mass of the hadronic system and
$Q$ is the hard scale of the process
(in the $B$ rest frame $Q=2E_X$, with $E_X$ denoting the hadronic energy).
In this regime, the dominant non-perturbative effects are 
 described by a single universal function, the {\it shape function},
entering the different spectra. At smaller $M_X^2\sim\Lambda_{QCD}^2$, however, few resonances dominate
the spectra and universality is lost.  It remains to be seen whether the
region where the shape function provides a good description of hadronic physics is large enough to allow a clean
extraction of the CKM parameters.

Several short-distance quantities based on various
partially integrated spectra or moments have been proposed
in the literature, with a particular attention to the problem of 
defining the shape function consistently beyond the
leading order~\cite{Akhoury:1995fp}.
Alternatively, one can use a model for the shape function, 
and estimate the theoretical uncertainty by changing the parameters of 
the model. This is routinely done  in order to implement cuts on
$M_X$ useful to identify the $B\to X_u l \nu$
sample~\cite{Dikeman:1997es}, an
approach  which 
has been actually used to extract $V_{ub}$ from the semileptonic
decay rate at LEP \cite{LEP}.
Very recently CLEO has also published a $V_{ub}$ measurement 
based on a combined analysis of the electron and photon 
spectra in semileptonic and radiative decays \cite{CLEO}. 
Other  proposals to identify
$B\to X_u l \nu$ events with low theoretical uncertainty 
include a cut on the invariant leptonic mass $q^2$
\cite{q2cut} and a combination of cuts on $q^2$ and $M_X$ \cite{combocut}.

In this paper, we propose a phenomenological strategy
to investigate  the ``shape-function region'' and possibly to
extract the ratio $\vert V_{ub}/(V_{tb} V_{ts}^*)\vert\approx \vert V_{ub}/V_{cb}|$. 
We use the fact that the differential rates 
$d\,\Gamma_{rd}/dx$  for the radiative decay $B\to X_s\gamma$  and 
$d\,\Gamma_{sl}/d\xi$ for the semileptonic decay $B\to X_u l \nu$ 
probe exactly the same long-distance structure function.
The kinematical variables involved are 
 $x=2E_\gamma/m_B$ and  $\xi=2 \,t/(1+t)$ with
 $t=\sqrt{1-M_X^2/E_X^2}$.
The ratio of the above two distributions  
is a short-distance quantity, which can be computed in perturbation theory.
As we will explain in section 4, to a good approximation it provides
a determination of $\vert V_{ub}/V_{cb} \vert$ in the SM. 
Including  perturbative $O(\alpha_s)$
corrections, one finds
\be\label{eq:sr}
 \left|\frac{V_{ub}}{V_{cb}}\right|^2\simeq
C(\alpha_s)
\frac{\frac{d}{d\xi}\Gamma_{sl}}{\left.\frac{d}{dx}\Gamma_{rd}\right|_{x=\xi} -
  \frac{\alpha_s}{\pi} B(\xi)}
\ee
where the coefficient $C(\alpha_s)$ and the function $B(\xi)$ will be
calculated in the following. Clearly, eq.\,(\ref{eq:sr})
is valid up to higher twist 
$O(1-x)\sim O(\Lambda_{QCD}/E_X)$ contributions.

Our treatment of the semileptonic and radiative decays
follows closely refs.~\cite{Aglietti:2000ub,Aglietti:2001cs,Aglietti:2001br}.
In this approach, the hard scale $Q$ is identified as twice the
hadronic energy $2E_X$ in the $B$ rest frame rather than as
the heavy-quark mass $m_b$. Consequently, 
the simplest kinematical variable for factorizing long-distance effects in this
decay turns out to be a function of $M_X/E_X$
and the relevant spectrum is $d\Gamma _{sl}/d\xi$ rather than
$d\Gamma _{sl}/dM_X$.

The advantages of our method are the following:
\begin{itemize}
\item the cancellation of Landau-pole effects and of non-perturbative
Fermi-motion effects in the ratio of spectra
 in eq.\,(\ref{eq:sr}) is exact and requires no approximation;

\item the shape function is defined in a consistent subtraction scheme
  for infrared (IR) singularities, avoiding double counting of IR logs;

\item  at least in principle, our ratio is valid point by point on
  the spectra, namely for any value of $x=\xi$.

\end{itemize}
In a realistic case, some smearing, such as partial
integration over the spectra, may be  necessary as all these calculations, 
including ours, rely on local quark--hadron duality.
One could integrate eq.\,(\ref{eq:sr})  over different kinematical ranges
and check the stability of the results with respect to the smearing
procedure. This, in turn, would amount to a test of the underlying
assumptions, and most notably of quark--hadron duality.

In this paper, we present a NLO calculation 
of the perturbative corrections to eq.\,(\ref{eq:sr}).
To make contact with experiments, we also give expressions of the
relevant functions in the presence of kinematical cuts on the hadronic
invariant mass $M_X$. 
After inclusion of the NLO corrections, the ratio  in eq.\,(\ref{eq:sr})
allows for a determination  of $|V_{ub}/V_{cb}|$ with a
$O(5\%)$ theoretical error.

The paper is organized as follows. In section \ref{sec:formalism} we 
introduce the formalism used throughout
the paper, while results for the relevant spectra are given in section \ref{sec:results} and in the Appendix.
The extraction of $|V_{ub}/V_{cb}|$ and the test of local duality 
are discussed in section \ref{sec:sumrule}. Section
\ref{sec:conclusions} contains our summary.

\section{Formalism}
\label{sec:formalism}
The heavy-quark Fermi motion in semi-inclusive 
decays~\cite{Ali:1979is,Altarelli:1982kh} can be treated in field
theory by introducing the shape function $\tilde f(k_+)$,
also known as the structure or light-cone distribution function of heavy
flavours \cite{Jaffe:1993ie,Bigi:1993ex,Neubert:1993ch,Neubert:1993um}.

The calculation of the spectra in these decays requires the Operator Product
Expansion (OPE) of the time-ordered product $T$ of two transition
operators  (currents or effective Hamiltonians). 
Schematically, this
involves the expansion of the propagator
\begin{eqnarray}
\label{eq:ope}
\frac{1}{M_X^2+2 E_X k_+ +k^2+i\varepsilon } \simeq
\frac{1}{M_X^2+i\varepsilon}\left[ \sum_n \left(\frac{-2 E_X
      k_+ }{M_X^2+i\varepsilon} \right)^n 
\right] + O\left(\frac{k^2}{M_X^2}\right)\,,
\end{eqnarray}
where $k_+=p_X\cdot k/E_X$ 
is the virtuality of the heavy quark ($p_b=m_B v+k$) and
$p_X$, $M_X$ and $E_X$ are momentum, mass and energy of the
final hadronic state, including the spectator quark\footnote{At the
leading-twist, the use of the meson mass $m_B$ instead of the
quark mass $m_b$ in the decomposition of heavy quark momentum $p_b$ only
shifts the integration range of $k_+$~\cite{Aglietti:1998mz}.}.
Since it is expected from the confinement dynamics
that $k_+\sim\Lambda_{QCD}$, 
one can identify three different kinematical regions,
depending on the value of $M_X^2$:
\begin{enumerate}
\item When $M_X^2\sim E_X^2\gg\Lambda_{QCD}^2$, the OPE converges. At
  leading twist accuracy, one can safely retain only 
  the lowest order ($n=0$) term in
  eq.\,(\ref{eq:ope}).

\item When $M_X^2\sim E_X \Lambda_{QCD}$, all the terms in the
  sum become of order 1. This is precisely the shape-function
  region. Indeed, the leading-twist
terms, i.e. those obtained by neglecting corrections of 
${ O}(k^2/M_X^2)\sim  { O}(
\Lambda_{QCD}/E_X)$, are resummed by introducing the shape
function, formally defined as
\begin{equation}
\tilde f(k_+)=\frac{\langle B\vert \bar h_v \delta(k_+-i D_+) h_v\vert B\rangle}{\langle B\vert \bar h_v h_v\vert B\rangle} \,,
\end{equation}
where $D_+$ is the light-cone plus-component of the covariant
derivative and $h_v$ is the heavy quark field
in the Heavy Quark Effective Theory;
$\tilde f(k_+)$ gives the probability of finding a heavy-quark with a
light-cone residual momentum $k_+$ inside the meson. 
Note that neglecting the terms of order $k^2/M_X^2$ in eq.\,(\ref{eq:ope})  is
equivalent to assuming the local quark--hadron duality. 
As the rate   is proportional to the imaginary part of eq.\,(\ref{eq:ope}),
it follows that the distribution in $k_+$ coincides with 
that in $-M_X^2/2E_X$ and is therefore experimentally accessible.

\item When $M_X^2\sim \Lambda_{QCD}^2$,
even the resummation fails and the contribution of single resonances to the spectra is not negligible. In this region,
terms of  ${ O}(\Lambda_{QCD}/E_X)$ are not negligible and
the shape function cannot be defined.
\end{enumerate}

A different framework to factorize long-distance effects is provided by the resummation theory in perturbative
QCD~\cite{Catani:1992ua}. The connection between the two approaches is discussed in refs.~\cite{Aglietti:2000ub,Aglietti:2001cs}.
In this framework, a factorized formula for a given distribution, for example the photon spectrum $d\Gamma_{rd}/dx$ in
$B\to X_s\gamma$, reads
\begin{equation}
\frac{d\Gamma_{rd}}{dx}= |V_{tb} V_{ts}^*|^2
\,\Gamma^0_{rd}\,
\left[K_{rd}(\alpha_s) f(x;\alpha_s) + D_{rd}(x;\alpha_s)\right]\,,
\label{eq:fact}
\end{equation}
where $V_{tb}$ and $ V_{ts}$ are the relevant CKM elements, 
$\Gamma^0_{rd}$ is a normalization factor, $x$ is a properly
chosen factorization variable, $x=2E_\gamma/m_B=1-M_X^2/m_B^2$ in this
example, while the
coefficient $K_{rd}$ and the remainder function $D_{rd}(x)$ are
short-distance quantities computable in perturbation theory.
The form factor $f(x)$ contains all the IR logarithms of the form
\begin{equation}\label{eq:logs}
 \alpha_s(2E_X)^k \left(\frac{\ln^n(1-x)}{1-x}\right)_+
\qquad\mbox{with }0\le n\le 2k-1\, ,
\end{equation}
where the plus-distribution is defined as
\begin{equation}
\left(\frac{\ln^n(1-x)}{1-x}\right)_+=\frac{\ln^n(1-x)}{1-x}-\delta(1-x
)\int_0^1 dy \frac{\ln^n(1-y)}{1-y}\,.
\end{equation}
The function $f(x)$ is well defined for
$1-x\gg \Lambda_{QCD}/m_B$.
The universality of $f(x)$ stems from the
general properties of the soft and the collinear radiation which produce the log-enhanced terms in
eq.\,(\ref{eq:logs}).
On the contrary, when $1-x\lsim \Lambda_{QCD}/m_B$, $f(x)$ is no longer calculable in
perturbation theory.  However, $f(x)$ is related to the shape function 
$\tilde f(k_+)$ in the following way~\cite{Aglietti:2001cs}
\begin{equation}
f(k_+)=\int dk_+^\prime C(k_+-k_+^\prime;\alpha_s)\tilde f(k_+^\prime) \,,
\end{equation}
where $k_+=-m_B(1-x)$ (in the case of $B\to X_s\gamma$, the hard scale
is $2E_X= m_B$).
The coefficient function $C(k_+-k_+^\prime;\alpha_s)$ remains perturbative also for $1-x\sim \Lambda_{QCD}/m_B$
and allows for a definition of a non-perturbative form factor in this region.

\section{Results}
\label{sec:results}
In this section we present a NLO determination of the short-distance 
functions entering the formulae of the photon energy 
spectrum in $B\to X_s\gamma$
and of the $z$-spectrum in $B\to X_u l \nu$, 
written in the factorized form of eq.\,(\ref{eq:fact}).
In the latter case, we also give expressions for the same functions in the presence of kinematical cuts.

\subsection{Photon spectrum in $B\to X_s\gamma$}
The photon spectrum in the radiative  $B\to X_s\gamma$ decay 
can be expressed as in eq.\,(\ref{eq:fact}).
The normalization factor $\Gamma _{rd}^{0}$ for the $b\to s\gamma$
rate is defined as
\begin{equation}
\label{eq:bsg}
\Gamma _{rd}^{0} \equiv \frac{\alpha}{\pi }\frac{G_{F}^{2}%
\,m_{b}^{3}\,m_{b,\overline{MS}}^{2}\left( m_{b}\right) }{32\pi ^{3}},
\end{equation}
where $\alpha=1/137.036$ is the fine structure constant.
The NLO relation of the $\overline{MS}$ mass,
 $m_{b,\overline{MS}}(m_b)=4.26\pm 0.09$
GeV~\cite{Lubicz:2000ch}, with the pole mass, $m_b\simeq 4.7$ GeV, 
both appearing in eq.\,(\ref{eq:bsg}), is
\begin{equation}
\frac{m_{b,\overline{MS}}\left( m_{b}\right) }{m_{b}}=1-\frac{\alpha
_{s}\left( m_{b}\right) C_{F}}{\pi }+O\left( \alpha_{s}^{2}\right) \,,
\end{equation}
where $C_F=(N^2-1)/(2N)$ and $N$ is the number of colours.

The coefficient function $K_{rd}$ reads
\begin{eqnarray}
\label{eq:kbsg}
&&K_{rd}\left( \alpha _{s} \right) =
C_{7}^{\left( 0\right) }\left( \mu _{b}\right) ^{2} \left[
1+\frac{\alpha
_{s}\left( \mu _{b}\right) }{2\pi }\sum_{i=1}^{8}\frac{C_{i}^{\left(
0\right) }\left( \mu _{b}\right) }{C_{7}^{\left( 0\right) }\left( \mu
_{b}\right) }\left(
{\rm Re}\,r_{i} +\gamma _{i7}^{\left( 0\right) }\ln
\frac{m_{b}}{\mu_{b}} \right)
\right.\\
&&~~ \hspace{4cm}\left.+\,\frac{\alpha_{s}\left( \mu _{b}\right) }{2\pi }
\frac{C_{7}^{\left( 1\right)}\left( \mu _{b}\right)}{C_{7}^{\left(
      0\right)}\left( \mu _{b}\right)}
+O\left( \alpha_{s}^{2}\right)
\right]
\,; \nn
\end{eqnarray}
the definitions of $r_i$  are collected in the Appendix.
We have denoted by $C_i^{(0)}$ and $C_i^{(1)}$  the
LO and NLO contributions to the {\it effective} Wilson coefficients of the
$\Delta B=1$ effective Hamiltonian ~\cite{Buras:1992tc,Chetyrkin:1996vx}.
The scale $\mu_b$ is a renormalization scale of $O(m_b)$
and $\gamma^{(0)}_{i7}$ are elements of the LO anomalous dimension matrix.
The terms proportional to $\ln(m_b/\mu_b)$ make the coefficient $K_{rd}$
renormalization scale-independent at the NLO. $K_{rd}$
depends on the top mass: for $M_t=174.3\pm 5.1$ GeV we find
\be
K_{rd}=0.1134 \left[1\pm 0.014\pm 0.053^{+0.001}_{-0.060} \right],\label{Krd}
\ee
 where the first two errors come from 
the top uncertainty and from the treatment of the charm quark mass in
the matrix elements containing charm loops. We have followed
\cite{Gambino:2001ew} and used $m_c/m_b=0.22\pm 0.04$ in $r_i$ 
(see Appendix).  
The central value of eq.\,(\ref{Krd}) has been obtained for $\mu_b=m_b$
and the last error by varying $\mu_b$ between $m_b/2$ and $2m_b$. 
We have  also included 
in $K_{rd}$ a power correction $-\lambda_2 \,(C_2^{(0)}-C_1^{(0)}/6) 
\,C_7^{(0)}/9 m_c^2$ related to 
long-distance contributions from $(c\bar{c})$ intermediate states
\cite{powermc}  and the electroweak effects 
following Ref.~\cite{ew}. We have adopted a light Higgs boson mass, $m_h$, as
suggested by global fits, but the sensitivity to $m_h$ is very small. 
The latter two contributions  modify $K_{rd}$  by about 
$+2.4\%$ and $-3.8\%$.

As is well known, NLO corrections have an important numerical impact
in radiative decays: they increase the inclusive branching ratio by
up to more than 30\%  \cite{Chetyrkin:1996vx}. In eq.\,(\ref{Krd}) the
effect of NLO corrections is about $(+20\pm 30)$\%, according to
the  value of $\mu_b$ between $m_b/2$ and  $2m_b$.
Recently, it has been shown \cite{Gambino:2001ew} 
that the dominant reason for the enhancement of the inclusive BR 
is the running of the $b$ quark
mass, which affects differently the  light quarks and
the top loops. Following this observation, it is  possible to 
stabilize  the
perturbative expansion for $B\to X_s \gamma$ by a rearrangement of 
higher-order effects. 
In this approach, NLO corrections to $K_{rd}$ are not
smaller, but have a different sign,
\be
K_{rd}=0.1448 \,(1 -0.99 \,\alpha_s -0.013)= 0.1119 \left[1 \pm 0.014 \pm
0.053^{+0.044}_{-0.088}\right].
\label{opt}
\ee
Here the second correction is due to electroweak and $1/m_c^2$
effects and the errors have the same meaning as in the
previous equation.
Notice that now the NLO correction is predominantly due to the matrix
elements of the effective operators.
For this reason, it is numerically close to the NLO corrections to the
coefficient in the semileptonic decay, as we will see later on.
In particular, the large scale dependence observed in eq.~(\ref{opt})
is mostly due to the large $O(\alpha_s)$ corrections to the matrix
elements and is dominated by the scale dependence of $\alpha_s(\mu_b)$
and not by the scale dependence of the Wilson coefficients,
in contrast to the situation in the preceding equation. In the following
numerical studies, we will employ eq.~(\ref{opt}).

We stress that the definition of the coefficient function
in  eq.\,(\ref{eq:kbsg}) and of the remainder function given below
corresponds to a particular factorization 
scheme. Specifically, following the convention
of \cite{Aglietti:2001br}, we absorb in $f(x)$ {\it only}
the IR logarithms of eq.\,(\ref{eq:logs}).

The remainder function is given by (the functions
$d_{ij}(x)$  are defined in the Appendix)
\begin{equation}
\label{eq:rbsg}
D_{rd}\left( x;\alpha _s\right) =\frac{\alpha_{s}\left( \mu _{b}\right) }{%
\pi }\sum_{i\leq j}^{1,8}
C_{i}^{\left( 0\right) }\left( \mu _{b}\right)
\,C_{j}^{\left( 0\right) }\left( \mu _{b}\right) 
\,d_{ij}\left( x\right)
+O\left( \alpha _{s}^{2}\right) .
\end{equation}

Let us comment on the behaviour of the remainder function at the end points
$x=0$ and $1$. The functions $d_{77}\left(x\right) $ and 
$d_{88}\left( x\right) $ have a
logarithmic singularity $\ln(1-x)$ for $x\to 1$, coming from power-suppressed
long-distance contributions.
They are not factorized into the function $f(x)$, which
contains only terms of the form $\ln^n(1-x)/(1-x)$.
At the lower end-point $x=0$, the functions $d_{77}\left( x\right) $ 
and $d_{78}\left( x\right) $ are finite,
while $d_{88}\left( x\right) $ has a power singularity 
$d_{88}\left( x\to 0\right) \sim 1/x$.
This limit corresponds to a soft (real) photon and the singularity 
would be regulated by including virtual photon
contributions so that $1/x\to (1/x) _+$. 
These contributions are not relevant to our analysis.
Finally, the function $d_{88}\left( x\right) $ has a logarithmic singularity
for $m_{s}\rightarrow 0$ because a non-zero photon energy regulates the soft
but not the collinear singularity. In this case, the strange mass has
to be retained (we adopt $m_s=m_b/50$). 
The function $D_{rd}(x)$ is shown in fig.~1 for
different values of $\mu_b$.

\begin{figure}[t]
\begin{center}
\begin{tabular}{cc}
\hspace{-0.5cm}\mbox{\epsfxsize=8.5cm\epsffile{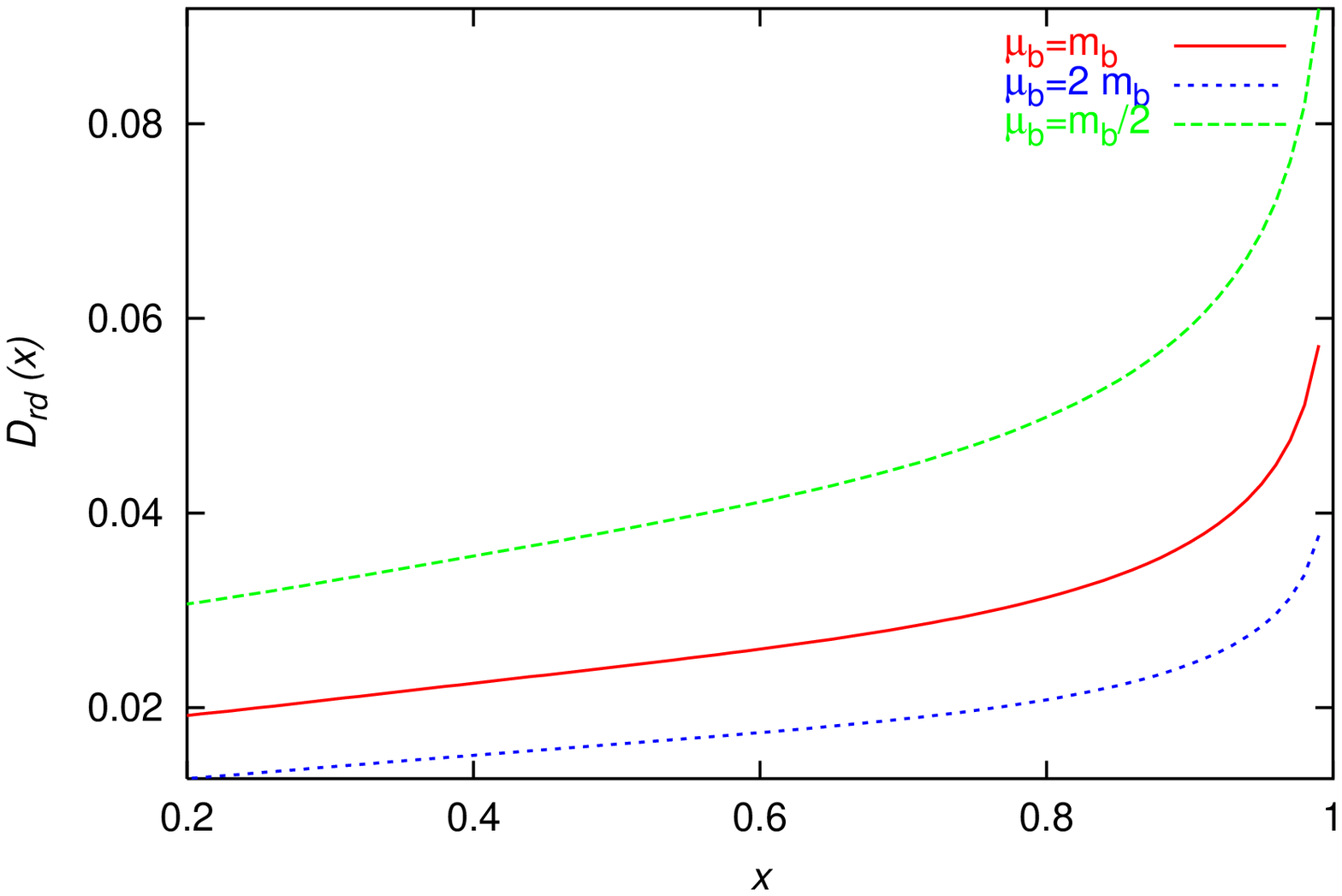}}&
\hspace{-0.5cm}\mbox{\epsfxsize=8.5cm\epsffile{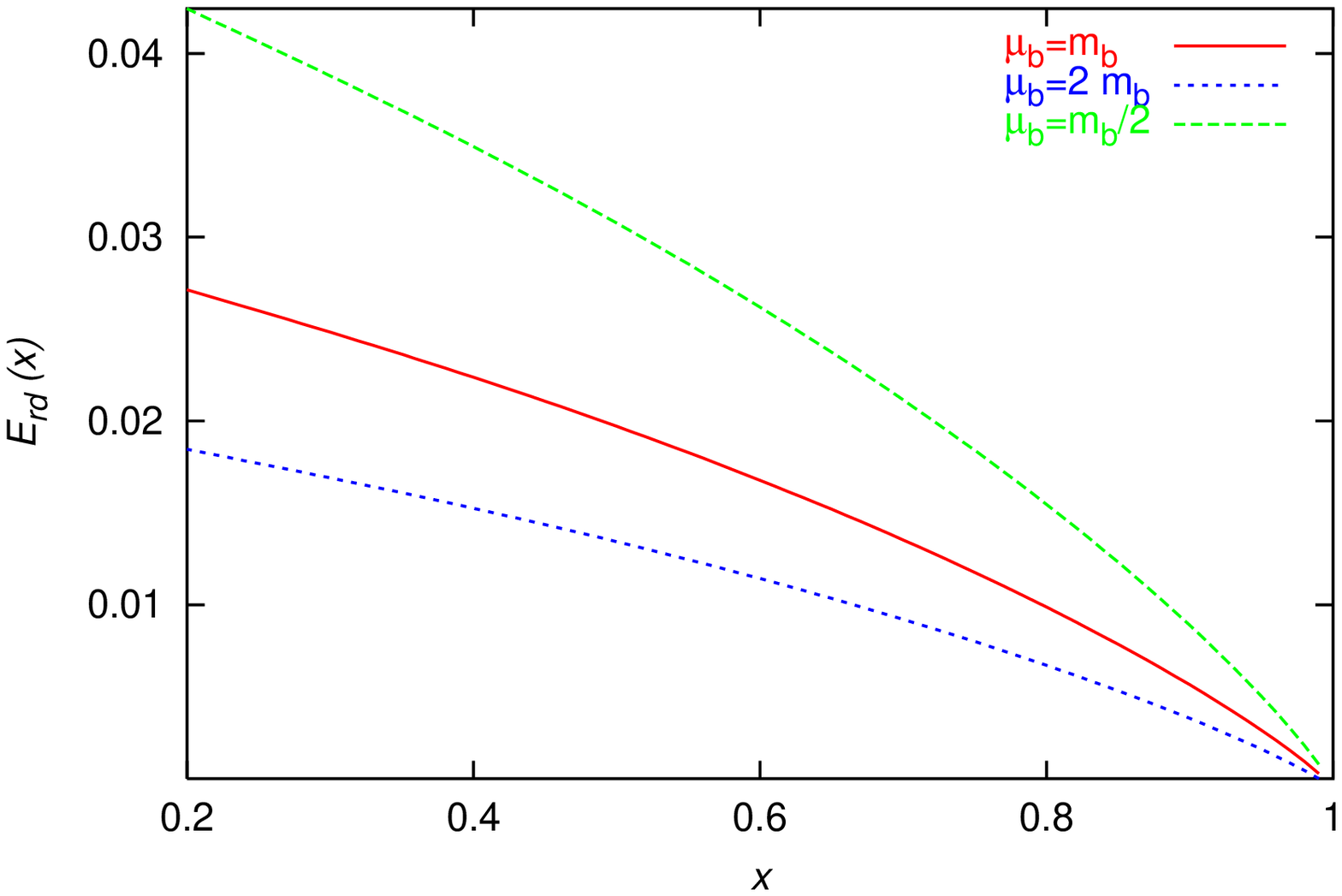}}
\end{tabular}
\end{center}
\caption{\sf 
The remainder functions $D_{rd}(x)$ and $E_{rd}(x)$
for $M_t=174.3$ GeV and different values of the renormalization scale
$\mu_b$.
}
\label{bsA}
\end{figure}

We also present results for the partially integrated rate, for which a
factorized formula also holds,
\begin{equation}\label{partint}
\Gamma_{rd}(x)\equiv\int_{x}^{1}\frac{d\Gamma _{rd}}{dx^{\prime
    }}dx^{\prime }=
 |V_{tb} V_{ts}^*|^2\,
\Gamma^0_{rd}\,\left[K_{rd}\left(
\alpha _{s}\right) \,F\left( x;\alpha _{s}\right) +E_{rd}\left( x;\alpha_{s}\right)\right]\, .
\end{equation}
The partially integrated form factor is defined as
\begin{equation}
F\left( x;\alpha _{s}\right) \equiv \int_{x}^{1}dx^{\prime }f\left(
x^{\prime };\alpha _{s}\right)\,,
\end{equation}
while the cumulative remainder function is given by
\begin{eqnarray}
E_{rd}\left( x;\alpha _{s}\right)\equiv \int_{x}^{1}dx^{\prime}D_{rd}(x^\prime)=
\frac{\alpha _{s}\left( \mu _{b}\right) }{
\pi }
\sum_{i\leq j}^{1,8}
C_{i}^{\left( 0\right) }\left( \mu _{b}\right)
\,C_{j}^{\left( 0\right) }\left( \mu _{b}\right)
f_{ij}\left( x\right)
\,+O\left( \alpha _{s}^{2}\left( \mu _{b}\right) \right)\, .\nn
\end{eqnarray}
Note that, contrary to the $d_{ij}$ case, the functions $f_{ij}$
vanish at the end-point $x=1$. 
The explicit expressions of the functions $f_{ij}(x)$ are 
given in the Appendix, and the cumulative remainder function 
is shown in fig.~1. Considering that  $K_{rd} \,F(x)\approx 0.1$ 
for  small $x$, the plot on the left side of fig.~1 shows that 
the scale dependence of the remainder function can be numerically
important. However,  the scale
dependence of $E_{rd}(x)$ is almost entirely due to the 
coupling constant $\alpha_s(\mu)$. 
Given the expression of the partially integrated rate in
eq.~(\ref{partint}), we observe a significant cancellation between the
$O(\alpha_s)$ correction in eq.~(\ref{opt}) and $E_{rd}(x)$. It
follows, in agreement with \cite{Gambino:2001ew},
that the NLO result for total rate is perturbatively quite stable.

\subsection{$\xi$ distribution in semileptonic decay}
The starting point for dealing with spectra in $B\to X_u l \nu$ is the 
triple-differential distribution, which  has been computed
at order $\alpha_s$ in ref.~\cite{DeFazio:1999sv}.
As demonstrated  in ref.~\cite{Aglietti:2001br}, the resummation 
of threshold logs takes a particularly simple form,
\begin{equation}
\frac{d^3\Gamma _{sl}}{dw dx_l dz}=
|V_{ub}|^2\,
\Gamma _{sl}^{0}\left[K_{sl}\left(w, x_l;\alpha
_{s}\right) \,f\left( z;\alpha _{s}\right) + D_{sl}\left(w, x_l, z;\alpha_{s}\right)\right]\,,
\label{triple}
\end{equation}
when one adopts  the kinematical variables 
\be
w=2 \frac{E_X}{m_B}, \ \ \ \ x_l=2 \frac{E_l}{m_B}, \ \ \ \ 
z=1-\frac{M_X^2}{4E_X^2}.
\ee
The main point is that all
IR logs are factorized in a function of  the single variable
$z$. Notice that $z$ is only defined between 3/4 and unity.
As already discussed, this formula can be generalized in the 
non-perturbative region corresponding to  $M_X^2\sim E_X \Lambda_{QCD}$,
where $f(z)$ is related
to the shape function defined in the effective theory
by a short-distance factor.
The $O(\alpha_s)$ expressions for $K_{sl}\left(w, x_l;\alpha_{s}\right)$ and
$D_{sl}\left(w, x_l, z;\alpha_{s}\right)$ can be found in 
ref.~\cite{Aglietti:2001br}. The integration over
$w$ and $x_l$ involves only short-distance functions and can
be easily carried
out~\cite{Aglietti:2001ub}. The resulting distribution reads
\begin{equation}
\label{sl}
\frac{d\Gamma _{sl}}{dz}=
|V_{ub}|^2\,
\Gamma _{sl}^{0}\left[K_{sl}\left( \alpha
_{s}\right) \,f\left( z;\alpha _{s}\right) +D^z_{sl}\left( z;\alpha
_{s}\right)\right]
\end{equation}
where
\begin{equation}
\Gamma _{sl}^{0}=\frac{G_{F}^{2}\,m_{b}^{3} \,m_{b,
 \overline{MS}}^2(m_b)}{192\pi ^{3}}.
\end{equation}
We have chosen a somewhat unusual  normalization 
factor $\Gamma _{sl}^{0}$, in analogy with the radiative decay case.
With our choice,  the  perturbative expansion 
of the total inclusive rate \cite{2loopincl} has a good convergence
 pattern. The coefficient is given by 
\begin{equation}
K_{sl}\left( \alpha _{s}\right) =1+\frac{\alpha _{s}C_{F}}{\pi }\left( \frac{%
403}{144}-\frac{\pi ^{2}}{2}\right) +O\left( \alpha_s^{2}\right) \simeq
1-0.91\,\alpha _{s}\,.
\end{equation}

Let us now consider a change of the factorization variable of the kind
\be
1-z\longrightarrow 1-z' =1-z + O(1-z)^2.
\label{change}
\ee
It simply amounts to a rearrangement of higher-twist contributions and
affects only the remainder function. Any ambiguity due to the choice
of variable is  related  to the underlying  leading-twist approximation and is
common to all leading-twist treatments of the shape
function\footnote{A first discussion of subleading twist effects 
in the shape function region has appeared recently \cite{subleading}. 
In $B\to X_s \gamma$ they seem to be small for $E_\gamma<2.3$ GeV 
($x\lsim 0.87$).}. 
The new factorized differential rate in terms of $z'$ is 
\be
\frac{d\Gamma_{sl}}{dz'}= |V_{ub}|^2 \, \Gamma^0_{sl} \left[
K_{sl}(\alpha_s) f(z',\alpha_s) + 
D_{sl}^{z'}(z',\alpha_s)\right],
\ee
where
\be
D_{sl}^{z'}(z';\alpha_s)= \frac{dz}{dz'} D_{sl}(z;\alpha_s)+
\frac{\alpha_s C_F}{\pi}\left[
\frac{\ln (1-z')+\frac74}{1-z'}
-\frac{dz}{dz'}\frac{\ln (1-z)+\frac74}{1-z} 
\right].
\label{Dslzp}
\ee
Here we have left implicit the functional dependence of $z=z(z')$.
The differential distributions $d\Gamma_{sl}/dz$ and $d\Gamma_{sl}/dz'$
are two examples of observables which differ only by higher-twist 
effects. In principle, a comparison between observables of this type 
could provide some information on the size of higher-twist contributions.

As we are going to directly compare the semileptonic and radiative
decays, it is preferable  to choose the  factorization variable
for the two cases in a similar way. In $B\to X_s \gamma$ 
the variable $z=1-M_X^2/(4 E_X^2)$ can be expressed in terms
of the photon energy fraction  $x$ as 
$z= 1- \frac{1-x}{(2-x)^2}$. Since we have factorized IR effects in
the radiative decay in terms of a function $f(x)$, it seems natural to 
adopt here a variable $\xi$
such that $1-z=(1-\xi)/(2-\xi)^2$. Solving for $\xi$ we then find
\be
1-\xi= \frac{1-t}{1+t},
\ee
where we have introduced the variable ($0\leq t\leq 1$)
\begin{equation}
t\equiv \sqrt{1-4\left( 1-z\right) }=\sqrt{1-M_X^{2}/E_X^{2}}\,.
\label{tdef}
\end{equation}
\begin{figure}[t]
\begin{center}
\begin{tabular}{cc}
\hspace{-0.5cm}\mbox{\epsfxsize=8.5cm\epsffile{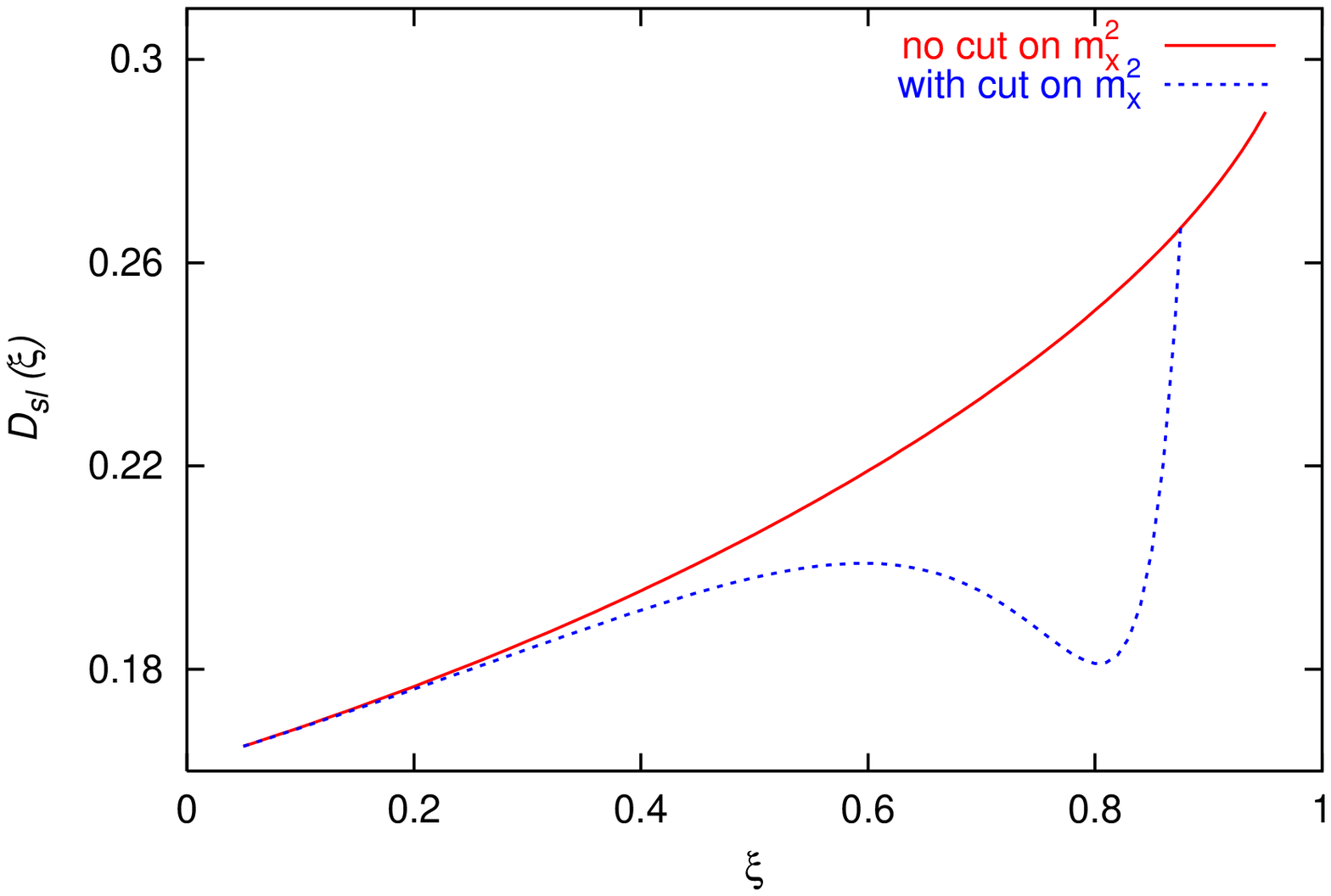}}
&\hspace{-0.5cm}\mbox{\epsfxsize=8.5cm\epsffile{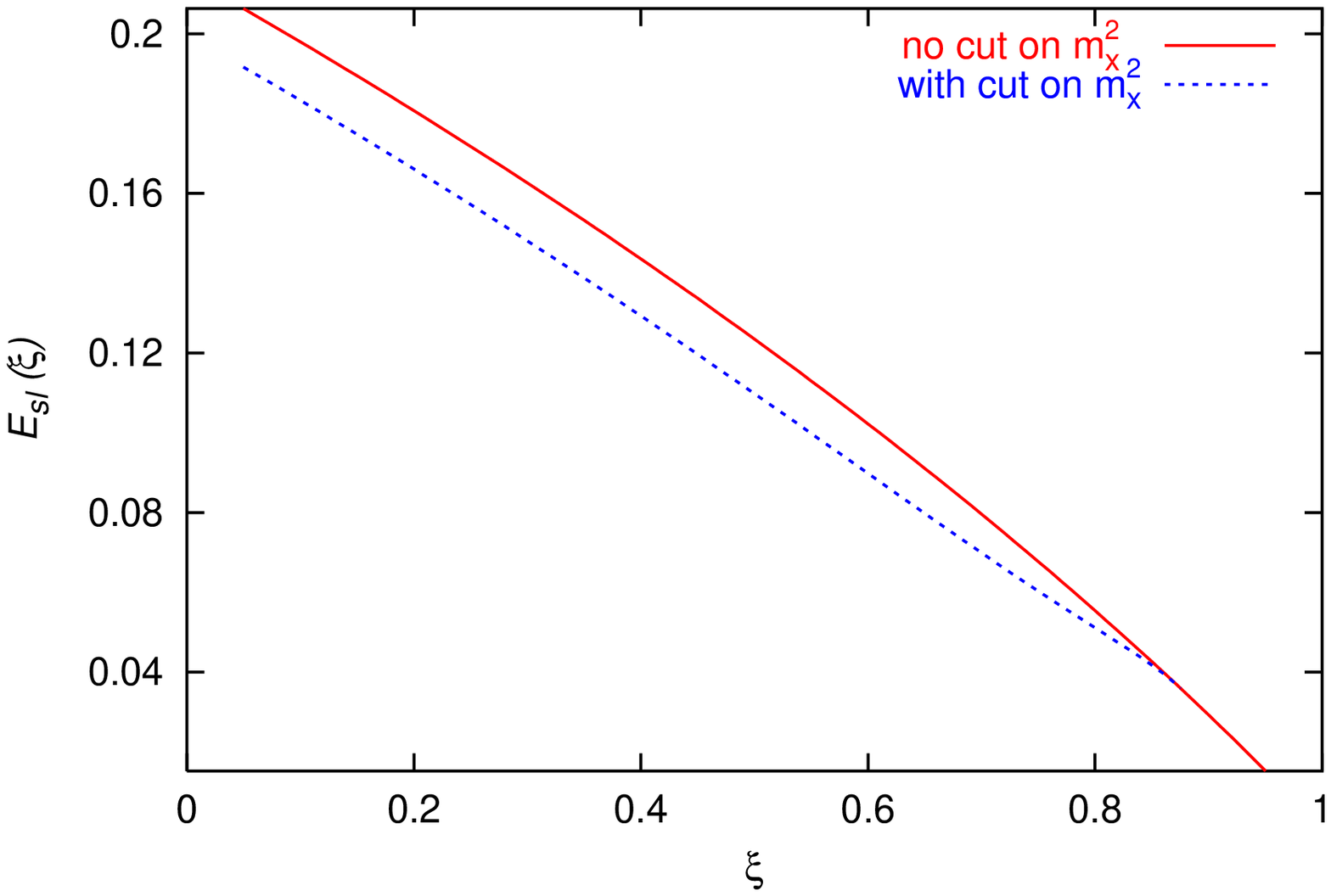}}
\end{tabular}
\end{center}
\caption{\sf
The remainder functions $D_{sl}(\xi)$ and $E_{sl}(\xi)$ 
with and without a cut on the invariant hadronic mass.}
\label{Dsl_zp}
\end{figure}
Unlike $z$, the variable $\xi$ is defined between 0 and 1.
Using $d z/d\xi= \xi/(2-\xi)^3$ and the formulae given in
\cite{Aglietti:2001br},
 it is straightforward to obtain $D_{sl}(\xi)\equiv D^\xi_{sl}(\xi)$, which is 
also shown in fig.~3:
\be
\label{eq:rfsldifxi}
D_{sl}\left( \xi; \alpha_s\right) =\frac{\alpha
_{s}C_{F}}{\pi } \left[
 \frac{21 + 21\,\xi - 5\,{\xi}^2 - {\xi}^3}{12} + 
   \frac{ 70 - 90\,\xi + 5\,{\xi}^2 + 8\,{\xi}^3 
      }{70} \ln (1 - \xi)\right].
\ee
The higher-twist logarithms contained in the remainder functions 
are process-dependent and in general cannot be factorized in the function
$f(\xi)$. It is nonetheless interesting to note that the IR logarithm
for $\xi\to 1$ has a very small coefficient in
eq.\,(\ref{eq:rfsldifxi}), much smaller than the corresponding
coefficient of $\ln (1-z)$ in  $D_{sl}^z(z)$.

One can also consider the partially integrated rate
\begin{equation}
\Gamma_{sl}(\xi)\equiv\int_{\xi}^{1}\frac{d\Gamma _{sl}}{d\xi^{\prime
    }} d\xi^{\prime }=|V_{ub}|^2 \,\Gamma^0_{sl}\left[K_{sl}\left(
\alpha_s\right) \,F\left( \xi;\alpha_s\right) +E_{sl}\left( 
\xi,\alpha_s\right)\right]\,,
\end{equation}
where at $O(\alpha_s)$ we find
\bea
E_{sl}(\xi;\alpha_s)&=& \frac{\alpha_{s}C_{F}}{\pi } \left[
  \frac{11725 - 6756\,\xi - 5898\,\xi^2 + 788\,\xi^3 +
    141\,\xi^4}{5040} \right.\nn\\
& & \left.
   \ \ \ \ \ \ \ \ \
+\frac{86 - 210\,\xi + 135\,\xi^2 - 5\,\xi^3 - 6\,\xi^4 
      }{210} \ln (1 - \xi)\right] .
\label{Eslxi}
\eea

If a cut on the hadronic mass $M_X< m_D$ is imposed, as often required in
a realistic environment, the remainder function in
eq.\,(\ref{eq:rfsldifxi}) is modified in the
region
$$
\xi<\xi_0\equiv1-c, \ \ \ \ \ \ c=\frac{m_D^2}{m_B^2}.
$$
The new expression is
\bea\label{Dslcut}
&&D_{sl}\left( \xi; \alpha_s\right)|_{\xi<\xi_0}=
\frac{\alpha_{s}C_{F}}{\pi }  \left\{
\frac{c^{\frac{3}{2}}\,\xi}{(1-\xi)^{\frac52}}\left[
  \frac{\left( 7 + 6\,c \right) \,\left(\xi-2 \right)
    \,\xi}{2}\right.\right.
\hfill \\
&&\hspace{1cm}- \left.
\frac{ 70\,{\left( 2 - \xi \right) }^2 + 
       10\,c^2\,\left( 1 - \xi \right)  + 
       7\,c\,\left( 35 - 35\,\xi + 8\,{\xi}^2 \right) 
     }{35} \ln (1 - \xi)
\right] \nn\\
&& \hspace{1cm}
 +\,\frac{21 - 42\,\xi + \left( 21 + 114\,c^2 + 16\,c^3 \right) \,
         {\xi}^2 - 2\,c^2\,\left( 57 + 8\,c \right) \,{\xi}^3 + 
        21\,c^2\,{\xi}^4  }{12\,{\left( 1 - \xi \right) }^3}
    \nn\\
&& \hspace{1cm}\left.
+\left[( 1 - \xi)^2 + c^3\,\xi
    \left( 2 - 3\xi + \xi^2 \right)  + 
   \frac{c^2\,\xi\left( 22 - 33\,\xi + 15\,\xi^2 - 2\,\xi^3 \right) }
    {2}\right]
\frac{\ln (1 - \xi)}{(1-\xi)^3}
\right\}.\nn
\eea
In the partially integrated case, the remainder function of 
eq.\,(\ref{Eslxi}) becomes
\be\label{Eslcut}
E_{sl}\vert_{ \xi<\xi_0} (\xi;\alpha_s)=E_{sl}\vert_{ \xi\ge \xi_0}
(\xi_0;\alpha_s) + \Delta(\xi_0;c) -\Delta(\xi;c)
\ee
for $\xi<\xi_0$, with 
\bea
\Delta(\xi;c)\hspace{-2mm}&\equiv&\int_{0}^\xi D_{sl} (\xi') \,d\xi'\nn\\
&=&
\frac{\alpha_s \, C_F}{\pi}\left\{c^2\,\left( 2 - \xi \right) \,\xi
   \frac{ 56\,c\,\left( 1 - \xi \right) +
        9\,\left( 22 - 22\,\xi + 3\,{\xi}^2 \right)   }{24\,
      {\left( 1 - \xi \right) }^2} 
- \frac{{\ln (1 - \xi)}^2}{2} \right. \\
&-& \hspace{-4mm}\left[\frac74-
   c^2\frac{  
        2c( 14 - 28\xi + 17{\xi}^2 - 3\,{\xi}^3)  + 
        3 \left( 33 - 66\xi + 44\,{\xi}^2 - 11{\xi}^3 + {\xi}^4
           \right)  }{6\,
      {\left( 1 - \xi \right) }^2}\right] \ln (1 - \xi)\nn\\
&+& c^{\frac{3}{2}}\,\xi
 \frac{          360\,c^2\,\left(  \xi -1\right)  - 
           35\,\left( 96 - 96\,\xi + 29\,{\xi}^2 \right)  - 
           14\,c\,\left( 246 - 246\,\xi + 61\,{\xi}^2 \right)   }{315\,
         {\left( 1 - \xi \right) }^{\frac32}} \nn\\
&+& 2\,c^{\frac{3}{2}}\,\left(  \xi-2 \right)
 \left.         \frac{      70\,{\left( 2 - \xi \right) }^2 + 
           30\,c^2\,\left( 1 - \xi \right)  + 
           7\,c\,\left( 41 - 41\,\xi + 8\,{\xi}^2 \right) 
         }{105\,{\left( 1 - \xi \right) }^{\frac32}}          
 \ln (1 - \xi)\nn
\right\}.
\eea
More complicated experimental cuts can be applied taking 
as a starting point the triple  differential distribution
\cite{Aglietti:2001br} and employing eq.\,(\ref{Dslzp}).

\section{Test of local duality and determination of $|V_{ub}/V_{cb}|$}
\label{sec:sumrule}
In this section we construct
a quantity that has a perturbative expansion in powers of $\alpha_s$
free from IR logs. In the shape-function region, this also guarantees
the cancellation of non-perturbative
effects related to the Fermi motion. 
Using the universality of the function $f(x)$, it follows from
eqs.~(\ref{eq:fact}) and (\ref{sl}) that
\be
\label{first}
\frac{1}{K_{sl}(\alpha_s)}\left[
\frac{1}{|V_{ub}|^2\,\Gamma^0_{sl}}\frac{d\Gamma_{sl}}{d\xi} -
D_{sl}(\xi;\alpha_s)\right]=
\frac{1}{K_{rd}(\alpha_s)}
\left[ \frac{1}{|V_{tb} V_{ts}^*|^2\,\Gamma^0_{rd}}
\frac{d\Gamma_{rd}}{dx} -  D_{rd}(x;\alpha_s)\right]_{x=\xi}.
\ee
Normalizing the experimental differential rates to the total semileptonic 
rate for $B\to X_c l \nu$, the previous equation can be rewritten as
\be
\left.
R\equiv
{\frac{ \frac{d }{d\xi}\BR_{sl,u} - 
\left|\frac{V_{ub}}{V_{cb}}\right|^2 
\frac{\BR_{sl,c}}{g_{sl}}\, D_{sl}(\xi;\alpha_s)}{
\frac{d }{dx}\BR_{rd} - \frac{6\alpha}{\pi} \left|
 \frac{V_{tb} V_{ts}^*}{V_{cb}}\right|^2 \frac{\BR_{sl,c}}{g_{sl}}
\,D_{rd}(x;\alpha_s)}}\right|_{x=\xi}
=\frac{K_{sl}(\alpha_s)}{K_{rd}(\alpha_s)} \frac{\pi}{6\alpha}
\left|\frac{V_{ub}}{V_{tb} V_{ts}^*}\right|^2
\label{new},
\ee
where $\BR_{sl,u}$ and $\BR_{rd}$ are
   the branching ratios of the semileptonic and radiative decay, respectively.
$\BR_{sl,c}$ denotes the experimental value for the
branching ratio of $B\to X_c l \nu$, $\BR_{sl,c}
\approx \BR_{sl,tot}=0.1045\pm 0.0021$ \cite{PDG}, while 
  $g_{sl}=g(m_c^2/m_b^2)$ is the phase-space function for the
semileptonic $b\to c$ decay, with $g(x)= 1-8x+8x^3-x^4 - 12 x^2 \ln
x$. Although it is sufficient to employ the tree-level expression for $g_{sl}$
in the above equations, there is an important  uncertainty from
 quark masses. A very accurate determination of this phase-space
factor, which could be adopted above, is $g_{sl}=0.575\pm
0.017$ \cite{Gambino:2001ew}.
The  CKM matrix elements appearing in eq.\,(\ref{new})
can be expressed in terms of the Wolfenstein parameters
$\lambda=|V_{us}|\simeq 0.22$, $\bar{\rho}$, and $\bar{\eta}$ as
\be
\label{ab}
\left|\frac{V_{ub}}{V_{cb}}\right|^2 = \lambda
\sqrt{\bar{\rho}^2+\bar{\eta}^2} + O(\lambda^3),\quad
\left|\frac{V_{ts}^*V_{tb}}{V_{cb}}\right|^2=
1+\lambda^2 (2\bar{\rho}-1) +O(\lambda^4)\approx
0.97. \nn
\ee

The r.h.s.\ of eq.\,(\ref{new})
is a purely short-distance quantity calculable in perturbation theory and
is independent  of the point $x=\xi$ where the l.h.s.\ is
evaluated. Hence one expects to find  an experimental plateau in the
range of $\xi$ where the shape function provides a good description of the
spectra. In practice, this allows a study of the extension of the
shape-function  region and a  test of  the validity
of the local quark--hadron duality hypothesis.

The numerical value of $R$ depends on the CKM matrix elements.
Let us introduce 
\be
C(\alpha_s)= \frac{6\alpha}{\pi} \frac{K_{rd}(\alpha_s)}{K_{sl}(\alpha_s)}\simeq
(1.94\pm 0.16)\times 10^{-3},
\label{Cc}
\ee
where we have  estimated  the theoretical error due to higher-order 
QCD corrections and to the top and charm masses on the basis of
the results of the previous section, combining different
uncertainties in quadrature. The ratio $R$ is then given by 
\bea
R &=&
\frac1{C(\alpha_s)}
\,\left|\frac{V_{ub}}{V_{tb}V_{ts}^{\ast }}
\right|^2
\label{eq:rval}
= 5.15 \times 10^2 \,(1\pm 0.08  )
\left|\frac{V_{ub}}{V_{tb}V_{ts}^{\ast }}
\right|^2 
\,,
\eea
Of course, the theoretical error on $R$ is not the only perturbative
uncertainty affecting eq.\,(\ref{new}), as higher-order corrections 
to the remainder functions $D_{sl}$ and $D_{rd}$ might not be
negligible, although they largely  cancel against each other in the ratio $R$.
In particular, we have seen that the remainder functions are relatively
sizeable. The error on the r.h.s.\ of eq.\,(\ref{new}) 
induced in this way depends on the precise shape of the spectra and
cannot easily be estimated.

Provided  a plateau can be experimentally identified, 
eq.\,(\ref{new}) provides a stringent constraint on the 
parameters of the unitarity triangle. 
In fact,  it allows for a model-independent extraction of
the ratio of  CKM matrix elements $\vert V_{ub}/V_{cb}\vert$.
This is apparent from the second of eq.~(\ref{ab}), which shows that
$|V_{tb} V_{ts}^*/V_{cb}|^2$ depends only weakly on
$\bar{\rho}$ and $\bar{\eta}$ (whose numerical values are taken from a
recent global fit \cite{CKMfit}).
Therefore, to very good approximation,
a measurement of the two spectra represents a measurement of $\vert
V_{ub}/V_{cb}\vert$. The master formula is
\be
\left|\frac{V_{ub}}{V_{cb}}\right|^2 \simeq 
C(\alpha_s) \frac{  \frac{d}{d\xi}  \BR_{sl,u}}{
\left. \frac{d}{dx}  \BR_{rd} \right|_{x=\xi} - h(\xi;\alpha_s)}\simeq
\left. C(\alpha_s) \frac{  \frac{d}{d\xi}  \BR_{sl,u}}{
 \frac{d}{dx}  \BR_{rd}} \right|_{x=\xi},
\label{master}
\ee
where $C(\alpha_s)$ has been defined in eq.\,(\ref{Cc})
and
\be\label{hxi}
h(\xi;\alpha_s)= \frac{6\,\alpha}{\pi} \frac{\BR_{sl,c}}{  g_{sl}} \left[
D_{rd}(\xi;\alpha_s)- \frac{K_{rd}}{K_{sl}} D_{sl}(\xi;\alpha_s)\right].
\ee
Whenever kinematical cuts  have to be applied, the remainder functions
have to be modified accordingly. In the case of a cut on the invariant
hadronic mass, the relevant formulae are given  in the previous 
section.

We plot the function $h(\xi)$ in fig.~3. Quite remarkably, 
the two remainder functions cancel each other almost completely
in the combination $h(\xi)$. In the range, $\xi\gsim 0.77$, corresponding to
the photon spectrum measured by CLEO \cite{cleophoton}, the function 
$h(\xi)$ is small with respect to the measured 
spectrum. Higher order corrections to the remainder functions 
could partially disrupt the cancellation only if they were very different 
in the matrix elements of the semileptonic
and radiative decays. An additional source of uncertainty is the
factor $K_{rd}/K_{sl}$ in eq.\,(\ref{hxi}). The effect of the error on
this factor given  in eq.~(\ref{Cc})   is also shown in fig.~3.
Although it is difficult to assess the impact of higher-order
corrections on $h(\xi)$, 
we can certainly neglect it in the final expression
as a first approximation. These considerations are fairly independent of
the presence of a cut on the hadronic invariant mass.
Of course, it is straightforward to include in the analysis 
the small correction due to the fact that $|V_{tb} V_{ts}^*/V_{cb}|$ is
not exactly one. 
\begin{figure}[t]
\begin{center}
\begin{tabular}{cc}
\hspace{-0.5cm}\mbox{\epsfxsize=8.5cm\epsffile{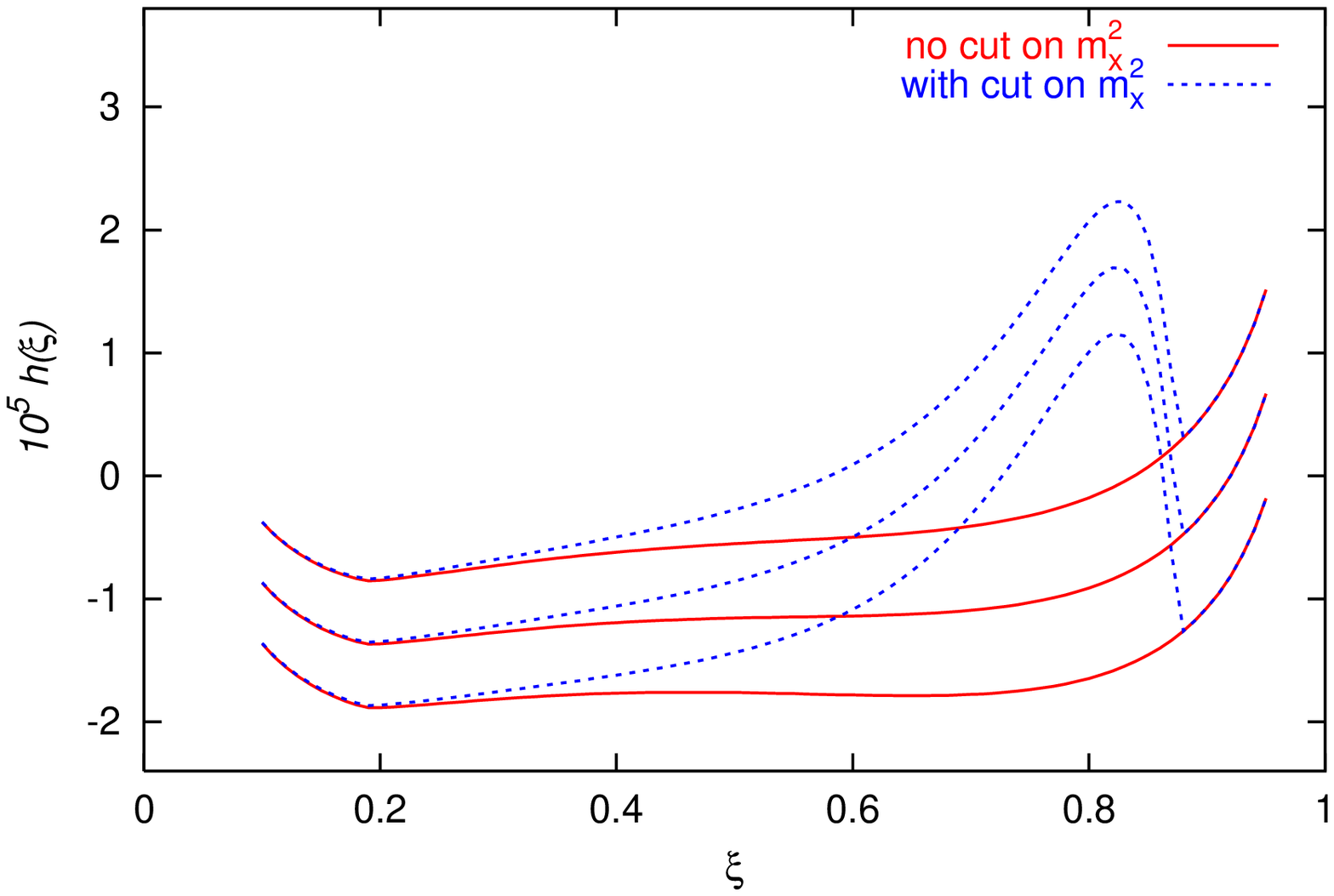}}
&\hspace{-0.5cm}\mbox{\epsfxsize=8.5cm\epsffile{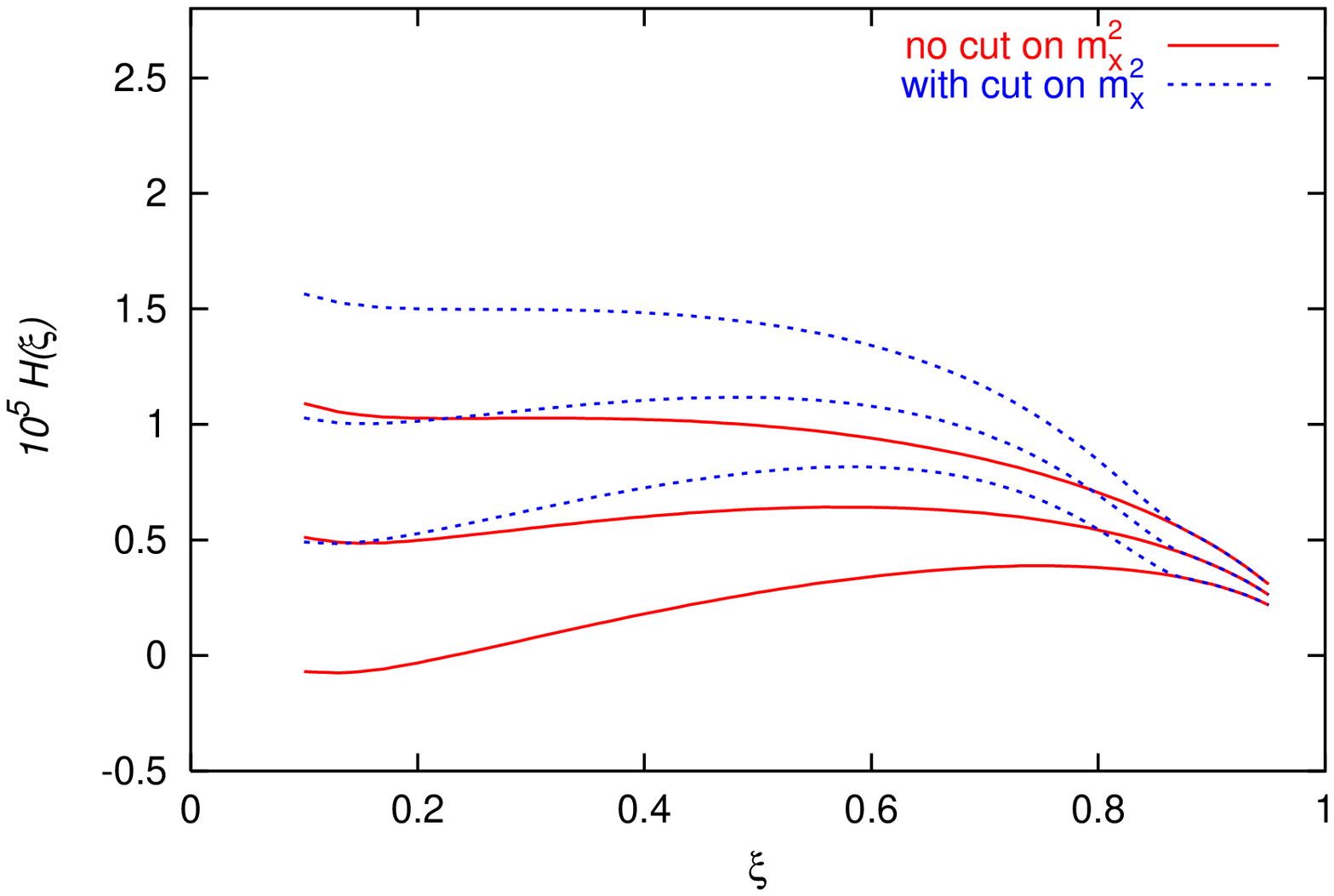}}
\end{tabular}
\end{center}
\caption{\sf 
The  functions $h(\xi)$ and $H(\xi)$ with and without a cut on 
the invariant hadronic mass ($\mu_b=m_b$). For each case, 
the three lines correspond to values of $K_{rd}/K_{sl}$ equal to our
central value $\pm 8\%$, see eq.~(\ref{Cc}).
}
\label{H_zp}
\end{figure}

In a realistic analysis, it may  be  necessary to introduce some
smearing  over the spectra before applying eq.\,(\ref{master}).
Some smearing can be provided by partial integration over the
spectra.  In this case our ratio takes the form
\begin{equation}
\label{eq:sri}
\left.
\frac{ {\BR}_{sl,u}(\xi) - \left|\frac{V_{ub}}{V_{cb}}\right|^2 
\frac{\BR_{sl,c}}{g_{sl}}\, E_{sl}(\xi)}{
\BR_{rd}(x) - \left|\frac{V_{tb} V_{ts}^*}{V_{cb}}\right|^2
\frac{\BR_{sl,c}}{g_{sl}} \,E_{rd}(x)}\right|_{x=\xi}
=R,
\end{equation}
where $\BR_{sl,u}(\xi)$ and $\BR_{rd}(x)$ are the partially integrated rates. 
The value of $R$  is given in 
eq.\,(\ref{eq:rval}). In analogy to the above discussion, one can
vary the value of $x=\xi$ where eq.\,(\ref{eq:sri})
 is evaluated and look for 
the presence of a plateau. 
The master formula of eq.\,(\ref{master}) then becomes
\be\label{master2}
\left|\frac{V_{ub}}{V_{cb}}\right|^2 \simeq 
\left.C(\alpha_s) \frac{  {\rm BR}_{sl,u}(\xi)}{
  {\rm BR}_{rd}(x) - H(\xi;\alpha_s)} \right|_{x=\xi}\simeq
\left.C(\alpha_s) \frac{  {\rm BR}_{sl,u}(\xi)}{
  {\rm BR}_{rd}(x) }\right|_{x=\xi}
\ee
where again we have neglected 
\be
H(\xi;\alpha_s)= \frac{6\,\alpha}{\pi} \frac{\BR_{sl,c}}{  g_{sl}} \left[
E_{rd}(\xi;\alpha_s)- \frac{K_{rd}}{K_{sl}} E_{sl}(\xi;\alpha_s)\right].
\ee
Indeed, as can be seen in fig.~3, $H(\xi,\alpha_s)$ is much smaller than
 the partially integrated branching fraction for $B\to
X_s \gamma$ (we recall that BR$_{rd}(0.77)\approx 3\times 10^{-4}$ 
\cite{cleophoton}). Although the uncertainty in the determination of
$H(\xi)$ is large in relative terms, it seems unlikely that it adds much to
the total uncertainty in the extraction of $|V_{ub}/V_{cb}|$, which is
therefore dominated by the error on $C(\alpha_s)$.

We stress that, since relation (\ref{first}) is valid
point by point, it is also possible to take into account 
a non-uniform  experimental efficiency in a model independent way,
provided that they are similar for the two rates.
For example, in the ideal case of a common efficiency $\epsilon(x=\xi)$,
eq.\,(\ref{master2}) will involve $\int dx' \,\epsilon(x') \,d\BR_{rd,sl}/dx'$.
In the presence of different efficiencies for the semileptonic and the 
radiative decays,
eq.\,(\ref{master2}) cannot be used in its present form 
and alternative startegies should be envisaged,
using for instance the same equation translated in the $N$-moment space
or deconvoluting the efficiencies.

\section{Conclusions}
\label{sec:conclusions}
We have constructed 
a ratio of spectra of semileptonic and radiative $B$ decays, from which
non-perturbative Fermi-motion effects completely cancel out. 
This is a short-distance quantity with an $\alpha_s$ expansion
not involving any IR logs. 
In spite of the semi-inclusive character of
the spectra involved, the expansion is similar
to those of inclusive quantities, such as  the total 
semileptonic width. Combined with the experimental determination of the
relevant spectra, our calculation allows for a test of the theory of the 
shape function and of the local quark--hadron duality
hypothesis.  
If a kinematical ``shape-function region'', where the shape function
describes well the observed spectra, exists,  the ratio $R$ introduced
in eq.\,(\ref{new}) should be independent of the variable $x=\xi$.
The quantity $R$ involves the photon spectrum of radiative decays
and the differential semileptonic charmless rate 
with respect to a certain function $\xi$ of the
hadronic mass over energy, $M_X/E_X$.

If indeed $R$ is constant over a sufficiently extended kinematical region,
 our formalism permits a transparent
extraction of $\vert V_{ub}/V_{cb}\vert$.
This is illustrated by a  very simple formula, eq.\,(\ref{master}),
that  links this CKM combination to the above two spectra.
Thanks in part to a remarkable cancellation of NLO effects, 
eq.\,(\ref{master}) is a very clean probe of the CKM structure. 
As this relation is valid point by point,  
the effects of experimental cuts and of a non-uniform
experimental efficiency can in principle be  incorporated.
Assuming that the data confirm the hypothesis of small higher twist effects,
a determination of 
$\vert V_{ub}/V_{cb}\vert$ with a $\sim 5\%$  theoretical 
error can be eventually obtained 
without using any model for the shape function.

The program we have just outlined can already be carried out using CLEO
and LEP data, in parallel to the other $|V_{ub}/V_{cb}|$
analyses, based on the electron energy  and  the 
hadronic invariant mass spectra. Unfortunately, 
the available photon spectrum  \cite{cleophoton}  is provided in the lab
frame; this effective smearing  may hamper a precise
comparison with the $\xi$ distribution. Moreover, 
LEP has a small sample of events and 
a low signal to background ratio, and therefore the  $\xi$ distribution
cannot be measured very precisely. 
We hope that some of these problems will be overcome 
at the $B$ factories. Even with the present
limitations, however, our method could provide a welcome check of the
alternative techniques and we urge  our experimental colleagues
to implement it.

\section*{Acknowledgements}
We are grateful to Marco Battaglia and Riccardo Faccini for useful discussions.
M.C. thanks for the hospitality the TH Division of CERN, 
where part of this work was done. The work of P.G.\ is supported by a
Marie Curie Fellowship of the European Union.

\section*{Appendix}
In this appendix we give all the  relevant constants and functions
necessary for the calculation of the differential spectrum 
$d\Gamma_{rd}/dx$. The non-vanishing functions $d_{ij}(x)$ 
 are given by~\cite{Pott:1995if,Greub:1996tg}
\begin{eqnarray}
d_{11}\left( x\right)  &=&\frac{1}{36}d_{22}\left( x\right) \,,  \ \ \
\ \ \ \ \ 
d_{12}\left( x\right)  =-\frac{1}{3}d_{22}\left( x\right) \,,  \nn \\
d_{17}\left( x\right)  &=&-\frac{1}{6}d_{27}\left( x\right) \,,  \ \ \
       \ \ \ \                          \ 
d_{18}\left( x\right)  =-\frac{1}{6}d_{28}\left( x\right) \,,  \nn \\
d_{22}\left( x\right)  &=&\frac{16c}{27}\int_{0}^{x/c}dt\left( 1-c\,t\right)
\left| \frac{G\left( t\right) }{t}+\frac{1}{2}\right| ^{2}\,,  \nn \\
d_{27}\left( x\right)  &=&-\frac{8c^{2}}{9}\int_{0}^{x/c}dt\,{\rm Re}\left[
G\left( t\right) +\frac{t}{2}\right] \,,  \nn \\
d_{28}\left( x\right)  &=&-\frac{1}{3}d_{27}\left( x\right) \,,  \nn \\
d_{77}\left( x\right)  &=&\frac{1}{3}\left[ 7+x-2x^{2}-2\left( 1+x\right)
\ln \left( 1-x\right) \right] \,,  \nn \\
d_{78}\left( x\right)  &=&\frac{2}{9}\left[ 4+x^{2}+4\left( 1-x\right) \,%
\frac{\ln \left( 1-x\right) }{x}\right] \,,  \\
d_{88}\left( x\right)  &=&\frac{1}{27x}\left[ -8+8x-x^{2}-2x^{3}+2\left(
2-2x+x^{2}\right) \ln \frac{m_{b}^{2}\left( 1-x\right) }{m_{s}^{2}}\right]\, . \nn
\end{eqnarray}

The constants $r_{i}$ are given by
\begin{eqnarray}
r_{1} &=&-\frac{1}{6}r_{2}\,,  \ \ \ \ \ \nn 
{\rm Re}~r_{2} =-4.092 -12.78 \,(0.29-m_c/m_b),\\
r_{7} &=&-\frac{10}{3}-\frac{8}{9}\pi ^{2}\,,  \ \ \ \ 
{\rm Re}~r_{8} =\frac{4}{27}\left( 33-2\pi ^{2}\right).
\end{eqnarray}
The remaining $r_i$ with $i=3$-6 have been calculated recently 
\cite{misiak2002}, but their effect is very small because the
corresponding Wilson coefficients are small. 
They can therefore be safely neglected.
The constants $r_{1}$ and $r_{2}$ depend very sensitively 
on the precise value of the charm mass. Following the arguments given
in \cite{Gambino:2001ew}, we use  $m_c/m_b=0.22\pm 0.04$ in $r_{1,2}$
and obtain  Re~$r_2=-4.99\pm 0.50$. For completeness, we also give the
value of $\gamma^{(0)}_{i7}$ entering eq.~(\ref{eq:kbsg}),
\begin{equation}
\gamma^{(0)}_{i7}=\left(-\frac{208}{243}, \frac{416}{81}, -\frac{176}{81}, -\frac{152}{243}, -\frac{6272}{81},\frac{4624}{243},
\frac{32}{3}, -\frac{32}{9}\right)\,.
\end{equation}

The functions $f_{ij}\left( x\right)$ entering the
partially integrated spectrum read
\begin{eqnarray}
f_{11}\left( x\right)  &=&\frac{1}{36}f_{22}\left( x\right) \,, \ \ \
\ \ \ \ \ 
f_{12}\left( x\right)  =-\frac{1}{3}f_{22}\left( x\right) \,,  \nn \\
f_{17}\left( x\right)  &=&-\frac{1}{6}f_{27}\left( x\right) \,,  \ \ \
\ \ \ \ \ 
f_{18}\left( x\right)  =-\frac{1}{6}f_{28}\left( x\right) \,,  \nn \\
f_{22}\left( x\right)  &=&\frac{16c}{27}\left\{ \left( 1-x\right)
\int_{0}^{x/c}dt\left( 1-c\,t\right) \left| \frac{G\left( t\right) }{t}+%
\frac{1}{2}\right| ^{2}+
\int_{x/c}^{1/c}dt\left( 1-c\,t\right) ^{2}\left|
\frac{G\left( t\right) }{t}+\frac{1}{2}\right| ^{2}\right\} \,,  \nn \\
f_{27}\left( x\right)  &=&-\frac{8c^{2}}{9}\left\{ \left( 1-x\right)
\int_{0}^{x/c}dt\,{\rm Re}\left[ G\left( t\right) +\frac{t}{2}\right]
+
\int_{x/c}^{1/c}dt\left( 1-c\,t\right) {\rm Re}\left[ G\left( t\right) +%
\frac{t}{2}\right] \right\} \,,  \nn \\
f_{28}\left( x\right)  &=&-\frac{1}{3}f_{27}\left( x\right) \,,  \nn \\
f_{77}\left( x\right)  &=&\frac{1}{9}\left( 1-x\right) \left[
31+x-2x^{2}-3\left( 3+x\right) \ln \left( 1-x\right) \right] \,,  \\
f_{78}\left( x\right)  &=&\frac{2}{27}\left[ 12\,\mathrm{Li}_{2}\left(
x\right) -12\left( 1-x\right) \ln \left( 1-x\right) +25-2\pi ^{2}-24x-x^{3}%
\right] \,,  \nn \\
f_{88}\left( x\right)  &=&\frac{1}{81}\left\{ -6\ln \frac{m_{b}}{m_{s}}%
\left( 3-4x+x^{2}+4\ln x\right) +12\,\mathrm{Li}_{2}\left( x\right) +24\ln
x+\right.   \nn \\
&& -3\left( 1-x\right) \left( 3-x\right) \ln \left( 1-x\right)
+\left( 1-x\right) \left( 28-5x-2x^{2}\right) -2\pi ^{2}\Big\} .  \nn
\end{eqnarray}
The function $G\left( t\right) $ in the integrand of $f_{22}\left( x\right) $
and $f_{27}\left( x\right) $ is given by
\begin{equation}
G\left( t\right) =
\left\{ \begin{array}{ll}
-2\arctan ^{2}\sqrt{\frac{t}{ 4-t}}  & \mathrm{for}\quad t< 4\\
2\ln ^{2}\frac{\sqrt{t}+\sqrt{t-4}}{2}-
2\pi i\ln\frac{ \sqrt{t}+\sqrt{t-4}}{2}-\frac{\pi ^{2}}{2}
& \mathrm{for}\quad t\geq 4.
\end{array}
\right.
\end{equation}
The above integrals can be performed analytically with the substitutions $%
t=4\sin ^{2}y$ for $t\leq 4$ and $t=4\cosh ^{2}y$ for $t\geq 4$.
The resulting expressions are quite long and are not reported here.
For practical purposes, the compact integral representation given above
   is adequate.

The relation between the $d_{ij}$ and $f_{ij}$ functions is simply given by
\begin{equation}
d_{ij}\left( x\right) =-\frac{df_{ij}}{dx}\left( x\right)\, .
\end{equation}


\begin{thebibliography}{99}
\newcommand{\np}[3]{Nucl. Phys. B {\bf #1} (#2) #3}
\newcommand{\pl}[3]{Phys. Lett. B {\bf #1} (#2) #3}
\newcommand{\pr}[3]{Phys. Rev.  D {\bf #1} (#2) #3}
\newcommand{\prl}[3]{Phys. Rev. Lett. {\bf #1} (#2) #3}
\newcommand{\prp}[3]{Phys. Rept. {\bf #1} (#2) #3}
\newcommand{\zpc}[3]{Z. Phys. {\bf C#1} (#2) #3}

\bibitem{Bigi:1993ex}
I.~I.~Bigi, M.~A.~Shifman, N.~G.~Uraltsev and A.~I.~Vainshtein,
Phys.\ Rev.\ Lett.\  {\bf 71} (1993) 496
[arXiv:hep-ph/9304225] and
Int.\ J.\ Mod.\ Phys.\ A {\bf 9} (1994) 2467
[arXiv:hep-ph/9312359].


\bibitem{Neubert:1993um}
M.~Neubert,
Phys.\ Rev.\ D {\bf 49} (1994) 4623
[arXiv:hep-ph/9312311].

\bibitem{Akhoury:1995fp}
R.~Akhoury and I.~Z.~Rothstein,
Phys.\ Rev.\ D {\bf 54} (1996) 2349
[arXiv:hep-ph/9512303];
A.~K.~Leibovich and I.~Z.~Rothstein,
Phys.\ Rev.\ D {\bf 61} (2000) 074006
[arXiv:hep-ph/9907391];
A.~K.~Leibovich, I.~Low and I.~Z.~Rothstein,
Phys.\ Rev.\ D {\bf 61} (2000) 053006
[arXiv:hep-ph/9909404];
A.~K.~Leibovich, I.~Low and I.~Z.~Rothstein,
Phys.\ Lett.\ B {\bf 486} (2000) 86
[arXiv:hep-ph/0005124];
M.~Neubert,
Phys.\ Lett.\ B {\bf 513} (2001) 88
[arXiv:hep-ph/0104280];
A.~K.~Leibovich, I.~Low and I.~Z.~Rothstein,
Phys.\ Lett.\ B {\bf 513} (2001) 83
[arXiv:hep-ph/0105066].


\bibitem{Dikeman:1997es}
V.~D.~Barger, C.~S.~Kim and R.~J.~Phillips,
Phys.\ Lett.\ B {\bf 251} (1990) 629;
A.~F.~Falk, Z.~Ligeti and M.~B.~Wise,
Phys.\ Lett.\ B {\bf 406} (1997) 225
[arXiv:hep-ph/9705235];
I.~I.~Bigi, R.~D.~Dikeman and N.~Uraltsev,
Eur.\ Phys.\ J.\ C {\bf 4} (1998) 453
[arXiv:hep-ph/9706520].

\bibitem{LEP}
R. Barate {\it et al.} (ALEPH Coll.), Eur. Phys. J. C {\bf 6} (1999) 555;
M. Acciarri {\it et al.} (L3 Coll.), Phys. Lett.  B {\bf 436} (1998);
P. Abreu {\it et al.} (DELPHI Coll.),  Phys. Lett. B {\bf 478} (2000) 14;
G. Abbiendi {\it et al.} (OPAL Coll.)  Eur. Phys. J. C {\bf 21} (2001) 399.

\bibitem{CLEO}
A.~Bornheim  [CLEO Coll.],
arXiv:hep-ex/0202019.


\bibitem{q2cut}
C.~W.~Bauer, Z.~Ligeti and M.~E.~Luke,
Phys.\ Lett.\ B {\bf 479} (2000) 395
[arXiv:hep-ph/0002161].

\bibitem{combocut}
C.~W.~Bauer, Z.~Ligeti and M.~E.~Luke,
Phys.\ Rev.\ D {\bf 64} (2001) 113004
[arXiv:hep-ph/0107074].


\bibitem{Aglietti:2000ub}
U.~Aglietti,
arXiv:hep-ph/0010251.
\bibitem{Aglietti:2001cs}
U.~Aglietti,
Phys.\ Lett.\ B {\bf 515} (2001) 308
[arXiv:hep-ph/0103002].
\bibitem{Aglietti:2001br}
U.~Aglietti,
Nucl.\ Phys.\ B {\bf 610} (2001) 293
[arXiv:hep-ph/0104020].


\bibitem{Ali:1979is}
A.~Ali and E.~Pietarinen,
Nucl.\ Phys.\ B {\bf 154} (1979) 519.

\bibitem{Altarelli:1982kh}
G.~Altarelli, N.~Cabibbo, G.~Corb\`o, L.~Maiani and G.~Martinelli,
Nucl.\ Phys.\ B {\bf 208} (1982) 365.

\bibitem{Jaffe:1993ie}
R.~L.~Jaffe and L.~Randall,
Nucl.\ Phys.\ B {\bf 412} (1994) 79
[arXiv:hep-ph/9306201].


\bibitem{Neubert:1993ch}
M.~Neubert,
Phys.\ Rev.\ D {\bf 49} (1994) 3392
[arXiv:hep-ph/9311325].

\bibitem{Aglietti:1998mz}
U.~Aglietti, M.~Ciuchini, G.~Corb\`o, E.~Franco, G.~Martinelli and L.~Silvestrini,
Phys.\ Lett.\ B {\bf 432} (1998) 411
[arXiv:hep-ph/9804416].

\bibitem{Catani:1992ua}
S.~Catani, L.~Trentadue, G.~Turnock and B.~R.~Webber,
Nucl.\ Phys.\ B {\bf 407} (1993) 3.

\bibitem{Lubicz:2000ch}
V.~Lubicz,
Nucl.\ Phys.\ Proc.\ Suppl.\  {\bf 94} (2001) 116
[arXiv:hep-lat/0012003].

\bibitem{Buras:1992tc}
A.~J.~Buras, M.~Jamin, M.~E.~Lautenbacher and P.~H.~Weisz,
Nucl.\ Phys.\ B {\bf 400} (1993) 37
[arXiv:hep-ph/9211304];
M.~Ciuchini, E.~Franco, G.~Martinelli and L.~Reina,
Nucl.\ Phys.\ B {\bf 415} (1994) 403
[arXiv:hep-ph/9304257].
\bibitem{Chetyrkin:1996vx}
K.~Chetyrkin, M.~Misiak and M.~Munz,
Phys.\ Lett.\ B {\bf 400} (1997) 206
[Erratum-ibid.\ B {\bf 425} (1997) 414]
[arXiv:hep-ph/9612313] and refs.\ therein.

\bibitem{Gambino:2001ew}
P.~Gambino and M.~Misiak,
Nucl.\ Phys.\ B {\bf 611} (2001) 338
[arXiv:hep-ph/0104034].

\bibitem{powermc} 
  M.B.~Voloshin, \pl{397}{1997}{275};
  A.~Khodjamirian {\it et al.}, \pl{402}{1997}{167};
  Z.~Ligeti, L.~Randall and M.B.~Wise, \pl{402}{1997}{178};
  A.K.~Grant, A.G.~Morgan, S.~Nussinov and R.D.~Peccei, \pr{56}{1997}{3151};
  G.~Buchalla, G.~Isidori and S.J.~Rey, \np{511}{1998}{594}.

\bibitem{ew} 
P.~Gambino and U.~Haisch,
JHEP {\bf 0110} (2001) 020 
[arXiv:hep-ph/0109058] and JHEP {\bf 0009} (2000) 001
[arXiv:hep-ph/0007259].



\bibitem{DeFazio:1999sv}
F.~De Fazio and M.~Neubert,
JHEP {\bf 9906} (1999) 017
[arXiv:hep-ph/9905351].

\bibitem{Aglietti:2001ub}
U.~Aglietti,
arXiv:hep-ph/0105168, to appear in the Proceedings of
``XIII Convegno sulla Fisica al LEP (LEPTRE 2001)'', Rome (Italy), 18-20 April 2001.

\bibitem{2loopincl} T.~van~Ritbergen, \pl{454}{1999}{353}.

\bibitem{subleading}
C.~W.~Bauer, M.~E.~Luke and T.~Mannel,
arXiv:hep-ph/0102089.

\bibitem{PDG} The Particle Data Group, Eur. Phys. J. C {\bf 15} (2000) 1


\bibitem{CKMfit}
M.~Ciuchini {\it et al.},
JHEP {\bf 0107} (2001) 013
[arXiv:hep-ph/0012308].

\bibitem{cleophoton}
S.~Chen {\it et al.}, CLEO Coll., \prl{87}{2001}{251807}.

\bibitem{Pott:1995if}
N.~Pott,
Phys.\ Rev.\ D {\bf 54} (1996) 938
[arXiv:hep-ph/9512252].


\bibitem{Greub:1996tg}
C.~Greub, T.~Hurth and D.~Wyler,
Phys.\ Rev.\ D {\bf 54} (1996) 3350
[arXiv:hep-ph/9603404];
A.~J.~Buras, A.~Czarnecki, M.~Misiak and J.~Urban,
Nucl.\ Phys.\ B {\bf 611} (2001) 488
[arXiv:hep-ph/0105160].

\bibitem{misiak2002}
A.~J.~Buras, A.~Czarnecki, M.~Misiak and J.~Urban,
arXiv:hep-ph/0203135.

\end{thebibliography}
\end{document}